# Search for Rare Nuclear Decays with HPGe Detectors at the STELLA Facility of the LNGS

P. Belli[a], R. Bernabei[a,b], F. Cappella[c,d], R. Cerulli[e], F.A. Danevich[f], A. d'Angelo[c,d], S. d'Angelo[a,b], A. Di Marco[a], M.L. Di Vacri[e], A. Incicchitti[c,d], V.V. Kobychev[f], G.P. Kovtun[g], N.G. Kovtun[g], M. Laubenstein[e], S. Nisi[e], D.V. Poda[f], O.G. Polischuk[c,f], A.P. Shcherban[g], D.A. Solopikhin[g], J. Suhonen[h], A.V. Tolmachev[i], V.I. Tretyak[f], R.P. Yavetskiy[i]

[a]*INFN, Sezione di Roma "Tor Vergata", Rome, Italy*
[b]*Dipartimento di Fisica, Università di Roma "Tor Vergata", Rome, Italy*
[c]*INFN, Sezione di Roma "La Sapienza", Rome, Italy*
[d]*Dipartimento di Fisica, Università di Roma "La Sapienza", Rome, Italy*
[e]*INFN, Laboratori Nazionali del Gran Sasso, Assergi (AQ), Italy*
[f]*Institute for Nuclear Research, Kyiv, Ukraine*
[g]*Kharkiv Institute of Physics and Technology, Kharkiv, Ukraine*
[h]*Department of Physics, University of Jyväskylä, Finland*
[i]*Institute for Single Crystals, Kharkiv, Ukraine*

**Abstract.** Results on the search for rare nuclear decays with the ultra low background facility STELLA at the LNGS using gamma ray spectrometry are presented. In particular, the best $T_{1/2}$ limits were obtained for double beta processes in $^{96}$Ru and $^{104}$Ru. Several isotopes, which potentially decay through different 2β channels, including also possible resonant double electron captures, were investigated for the first time ($^{156}$Dy, $^{158}$Dy, $^{184}$Os, $^{192}$Os, $^{190}$Pt, $^{198}$Pt). Search for resonant absorption of solar $^7$Li axions in a LiF crystal gave the best limit for the mass of $^7$Li axions (< 8.6 keV). Rare alpha decay of $^{190}$Pt to the first excited level of $^{186}$Os ($E_{exc}$ = 137.2 keV) was observed for the first time.

**Keywords:** Double β decay; Double electron capture; Alpha decay; Axions; Ultra low background HPGe spectrometry.
**PACS:** 23.40.-s; 23.60.+e; 14.80.Va; 29.30.-h.

## INTRODUCTION

Double beta (2β) decay is a process of rare nuclear transformation (A,Z) → (A,Z±2) with simultaneous emission of two electrons or positrons. Two neutrino (2ν) double β decay, when two (anti)neutrinos are also emitted, is allowed in the Standard Model (SM) of particle physics, but, being the second order process in weak interactions, it is the rarest nuclear decay observed to-date, with half lives in the range of $T_{1/2} \sim 10^{18} - 10^{24}$ yr [1-3]. Two neutrino 2β transitions to the ground states of daughter nuclei are the most probable, but for $^{100}$Mo and $^{150}$Nd transitions to the first $0_1^+$ excited levels are also observed; see summary in recent Ref. [4]. Studies of 2β2ν decays to the ground states and excited levels of daughter nuclei allow to test different theoretical approaches in the calculation of nuclear matrix elements for 2β decay.

Neutrinoless (0ν) double β decay is forbidden in the SM because it violates the lepton number conservation; however, it is predicted by many SM extensions, which describe neutrino as Majorana particle (identical to its antiparticle) with non-zero mass. Only $T_{1/2}$ limits for 2β0ν decay are known to-date (except of a positive claim for $^{76}$Ge [5]), with the best values up to $T_{1/2} > 10^{24} - 10^{25}$ yr for few 2β candidates. Experimental searches for 2β0ν decay are considered today as one of the most important tasks in modern particle and nuclear physics, which is able to shed light on the lepton number conservation, nature of neutrino (is it a Majorana or a Dirac particle), absolute scale of ν masses, possible existence of right-handed currents in weak interactions and other important issues [1-3].

Ultra low background conditions, which are necessary to investigate rare 2β decays, allow also to look for other processes with extra low probabilities such as rare α and β decays or solar axions. The latter particle appears as consequence of the Peccei-Quinn solution of the so-called strong CP problem in quantum chromodynamics [6].

# EXPERIMENTAL MEASUREMENTS

All the measurements described here were performed in underground conditions of the Laboratori Nazionali del Gran Sasso (LNGS) of the INFN (Italy) on the depth of 3600 m w.e. Some HPGe (high purity Germanium) detectors of the STELLA (SubTErranean Low Level Assay) facility for ultra low background γ spectrometry [7] were used to search for various rare nuclear processes, which are accompanied by emission of gamma quanta. The underground location decreases the flux of cosmic muons by a factor of ~$10^6$, and the massive passive shielding made of selected materials (Cu, low radioactive Pb, Cd, polyethylene), together with the flushing of HPGe set-ups with Rn-free $N_2$ gas, is used to suppress the natural radioactivity from the walls and the air of the laboratory. The searches performed in 2011-2013 are summarized below.

## Search for 2β decays of $^{96}$Ru and $^{104}$Ru

$^{96}$Ru has a quite high natural abundance δ = 5.54% [8] and high energy release $Q_{2\beta}$ = 2714.50 ± 0.12 keV [9]. It is one of only six isotopes-candidates for $2\beta^+$ decay [10], however, its decay up to recent times was searched for in only one short (178 h) experiment in 1985 performed above ground with two HPGe detectors (110 cm$^3$ each) using a Ru sample of 50 g. The peaks related with $2\beta^+$ decay (as well as with electron capture accompanied by positron emission $\epsilon\beta^+$ or double electron capture 2ε) were not found in the accumulated spectrum, and only $T_{1/2}$ limits were set on the level of $10^{16}$ yr [11].

In our preliminary measurements [12] at LNGS with a HPGe detector of 468 cm$^3$ during 158 h with Ru sample of 473 g, the $T_{1/2}$ limits were improved up to $10^{18} - 10^{19}$ yr. However, it was found that the sample contained $^{40}$K (3.4 Bq/kg) and cosmogenic $^{106}$Ru (24 mBq/kg). Both of them contribute to the 511 keV peak expected also in $2\beta^+$ and $\epsilon\beta^+$ processes in $^{96}$Ru, and it was clear that longer measurements should be done only after Ru purification. Purification by electron beam melting decreased $^{40}$K to 150 mBq/kg, and also $^{106}$Ru decreased to 5 mBq/kg due to its decay ($T_{1/2}$ = 374 d). New measurements during 5479 h were performed with the purified Ru sample (720 g) using a set-up with four HPGe detectors (~225 cm$^3$ each) in one cryostat. However, even with higher experimental sensitivity, the $2\beta^+/\epsilon\beta^+/2\epsilon$ processes in $^{96}$Ru were not observed, and only $T_{1/2}$ limits were set on the level of $10^{20} - 10^{21}$ yr [13]. In particular, they are also better than those established in recent measurements in the HADES underground laboratory (500 m w.e.) performed with Ru sample of 149 g with two HPGe detectors (395 and 325 cm$^3$) during 2592 h ($T_{1/2} > 10^{19} - 10^{20}$ yr) [14].

In addition, a limit of $T_{1/2} > 6.5 \times 10^{20}$ yr at 90% C.L. was given for $2\beta^-$ decay of another Ru isotope, $^{104}$Ru (δ = 18.62%, $Q_{2\beta}$ = 1301.2 ± 2.7 keV) for decay to the first excited level of $^{104}$Pd with $E_{exc}$ = 556 keV [13].

## Search for 2β decays of $^{156}$Dy and $^{158}$Dy

Two Dy isotopes, present in natural isotopic mixture, are potentially 2β active: $^{156}$Dy with abundance δ = 0.056% and energy release $Q_{2\beta}$ = 2005.95 ± 0.10 keV (candidate for $\epsilon\beta^+/2\epsilon$ decays) and $^{158}$Dy (δ = 0.095%, $Q_{2\beta}$ = 282.7 ± 2.5 keV, candidate for 2ε capture). First $T_{1/2}$ limits for 2β processes in Dy were obtained with a $Dy_2O_3$ sample (322 g, purity grade 99.98%) using a HPGe detector of 244 cm$^3$ in measurements during 2512 h [15]. Slight contamination of the sample by U/Th chains and also by $^{176}$Lu (9 mBq/kg) was found. No peaks related to 2β processes in $^{156}$Dy and $^{158}$Dy were observed, and only limits were set in the range of $T_{1/2} > 1.8 \times 10^{14} - 7.1 \times 10^{16}$ yr.

$^{156}$Dy and $^{158}$Dy isotopes are also interesting because in both of them the so-called resonant neutrinoless double electron capture (r-2ε0ν) is possible. In this process, initial and final (excited) atoms are mass-degenerated, and in case of perfect energy coincidence between the released energy and the energy of an excited state (~ 10 eV), an enhancement of the decay rate by a few orders of magnitude could be expected (see [16] and Refs. therein). While many 2β isotopes were excluded from the list of perspective r-2ε0ν candidates after recent precise measurements of atomic masses, $^{156}$Dy is still in the list [17]. The energy release for $^{158}$Dy is known currently with accuracy of only 2.5 keV, and it would be interesting to re-measure it with better precision.

As by-product, limits for α decays of $^{156,158,160,161,162}$Dy to excited $^{152,154,156,157,158}$Gd were also established: $T_{1/2} > 10^{16} - 10^{17}$ yr [15].

## Search for 2β decays of $^{184}$Os and $^{192}$Os

Two natural Os isotopes are potentially 2β active: $^{184}$Os (δ = 0.02%, $Q_{2β}$ = 1450.9 ± 1.0 keV, candidate for $εβ^+/2ε$ decays) and $^{192}$Os (δ = 40.78%, $Q_{2β}$ = 408 ± 3 keV, candidate for $2β^-$ decay). Up to-date, quite poor limits on probabilities of 2β processes in these isotopes were known: $T_{1/2} > 10^{10} – 10^{13}$ yr, extracted from an old experiment performed with photoemulsion (see [10]). In our studies [18] with a HPGe of 468 cm$^3$, natural Os (173 g) of purity grade > 99.999% was used. It was found that the Os sample is very pure: the measured during 2741 h spectrum with Os practically has no excess in comparison with the background, and only some presence of cosmogenic $^{185}$Os ($T_{1/2}$ = 93.6 d, 3 mBq/kg), and also $^{137}$Cs (2 mBq/kg), $^{207}$Bi (0.4 mBq/kg) was observed. Only $T_{1/2}$ limits were established for 2β decays: $T_{1/2} > \sim 10^{16} – 10^{17}$ yr for $^{184}$Os, and $T_{1/2} > 5.3×10^{19}$ yr for $^{192}$Os [18]. The results for $^{184}$Os are restricted by its low natural abundance.

## Search for 2β decays of $^{190}$Pt and $^{198}$Pt

The $T_{1/2}$ limits for 2β processes in $^{190}$Pt (δ = 0.014%, $Q_{2β}$ = 1384 ± 6 keV, candidate for $εβ^+/2ε$ decays) and $^{198}$Pt (δ = 7.163%, $Q_{2β}$ = 1049.2 ± 2.1 keV, candidate for $2β^-$ decay) were obtained with a HPGe of 468 cm$^3$ and Pt sample of 42.5 g (in fact, it was two Pt cups borrowed in the LNGS chemical laboratory) [19]. Measurements during 1815 h showed that the Pt is polluted by $^{192m}$Ir (40 mBq/kg) and $^{137}$Cs (7 mBq/kg), but not polluted by $^{40}$K, $^{60}$Co, and U/Th chains (this information is important for growth of crystals in Pt crucibles). Limits for possible 2β transitions were established as: $T_{1/2} > 8.4×10^{14} – 3.1×10^{16}$ yr for $^{190}$Pt, and $T_{1/2} > 3.5×10^{18}$ yr for $^{198}$Pt (earlier results were absent or very poor, $\sim 10^{11}$ yr, see [10]). In particular, for resonant 2ε0ν capture in $^{190}$Pt to the excited level of $^{190}$Os with $E_{exc}$ = 1382 keV, the limit is $T_{1/2} > 2.9×10^{16}$ yr at 90% C.L. The atomic mass difference $^{190}$Pt – $^{190}$Os is known now with accuracy of only 6 keV. It could be one more interesting aim of precise Penning trap measurements [20].

## First observation of α decay $^{190}$Pt → $^{186}$Os$^*$ ($E_{exc}$ = 137.2 keV)

Additional aim of the measurements with the Pt sample was to search for α decay of $^{190}$Pt to excited levels of $^{186}$Os. Because of the exponential dependence of $T_{1/2}^α$ on energy release, the most probable is α decay to the first excited level with $E_{exc}$ = 137.2 keV. Theoretical estimations, performed in different approaches (see details in [21]), were in the range of $T_{1/2,th}^α = (3.2 – 7.0)×10^{13}$ yr, and it was clear that this effect has a good chance to be registered with HPGe detectors available notwithstanding the very low $^{190}$Pt abundance (δ = 0.014%). In the measurements described above, the effect was clearly observed at 7.7σ (area under the 137.2 keV peak was equal to 132 ± 17 counts), and the corresponding half life is $T_{1/2,exp}^α = (2.6^{+0.4}_{-0.3}(stat.) ± 0.6(syst.))×10^{14}$ yr [21].

In fact, all six naturally occurring isotopes of platinum are potentially unstable in relation to α decay. Limits for α decay of other Pt isotopes accompanied by γ quanta were also established; they are in the range of $T_{1/2} > 3.6×10^{15} – 6.8×10^{19}$ yr [21].

## Search for resonant absorption of $^7$Li solar axions in a LiF crystal

The axion is a hypothetical particle, which appears as result of solution of the "strong CP problem" by Peccei & Quinn, with further modifications by Kim-Shifman-Vainstein-Zakharov and Dine-Fischler-Srednicki-Zhitnitskii (see recent review [6]). If the axion exists, the Sun could be an intense source of these particles; in particular, they are emitted instead of γ quanta in magnetic transitions in deexcitations of excited nuclear level of $^7$Li ($E_{exc}$ = 477.6 keV) populated in the main pp cycle of the solar nuclear reactions: $^7$Be + $e^-$ → $^7$Li$^*$ + $ν_e$. Arriving to Earth from the Sun, these quasi-monoenergetic axions could excite the same nuclei ($^7$Li), and one could look for deexcitation γ quanta of 477.6 keV with a proper detector located near a Li sample (or with a detector that contains Li nuclei). The area under the 477.6 keV peak is related with the mass of hadronic axion as $S \sim m_a^4$.

Few Li-containing samples were used in our measurements. Preliminary investigations [22] showed that LiF powders are highly contaminated by U/Th chains while LiF(W) crystals are radioactively pure (U/Th < ~0.01 Bq/kg). Thus in the final studies [23] a LiF crystal (553 g) was measured with a HPGe detector of 244 cm$^3$ during 4044 h. The peak at 477.6 keV was absent in the spectrum accumulated; this allowed to set a limit of $m_a$ < 8.6 keV at 90% C.L., which is the best limit for the mass of solar $^7$Li axions.

It should be noted that:

(1) In the used approach the axions are coupled to nucleons both in the production and in the absorption processes, and thus the $m_a$ limit is related only to the axion-nucleon coupling constant $g_{aN}$; uncertainties related with the coupling constants to gamma quanta $g_{a\gamma}$ and to electrons $g_{ae}$ disappear;

(2) The obtained limit is related with the pp solar cycle, the main source of the solar energy, thus it is not subject to uncertainties in the determination of the elemental abundances in the solar core;

(3) Joining the determined limit with the results of similar experiments with $^{57}$Fe nuclei, one can extend a window in the excluded hadronic axion masses to the limits [477.6, 0.145] keV.

## CONCLUSIONS

Various $2\beta$ processes, including resonant $2\epsilon 0\nu$ captures, were searched for in $^{96,104}$Ru, $^{156,158}$Dy, $^{184,192}$Os, $^{190,198}$Pt isotopes by means of ultra low background HPGe spectrometry. Only $T_{1/2}$ limits were established in the range of $T_{1/2} > 1.8\times10^{14} - 1.0\times10^{21}$ yr. These values are mostly the best today, sometimes better than the previous ones by few orders of magnitude, and sometimes obtained for the first time.

It would be interesting to re-measure more precisely $Q_{2\beta}$ values for isotopes where resonant $2\epsilon 0\nu$ capture is possible and where $Q_{2\beta}$ are known with low accuracy: $^{190}$Pt – $Q_{2\beta} = 1384 \pm 6$ keV, $^{158}$Dy – $Q_{2\beta} = 282.7 \pm 2.5$ keV.

Resonant absorption of hypothetical hadronic solar $^7$Li axions in a LiF crystal was also looked for with the HPGe detector. The effect is not observed, the obtained limit on axion mass $m_a < 8.6$ keV is the best for $^7$Li axions.

Alpha decay $^{190}$Pt $\to$ $^{186}$Os$^*$ ($E_{exc} = 137.2$ keV) was observed for the first time; the measured probability is 0.25% that corresponds to $T_{1/2} = 2.6\times10^{14}$ yr.

Measurements with the Os sample are in progress now aimed to observe $\alpha$ decays of $^{184}$Os and $^{186}$Os to excited states of $^{180}$W and $^{182}$W, respectively.

## ACKNOWLEDGMENTS

The group from the Institute for Nuclear Research (Kyiv, Ukraine) was supported in part by the Space Research Program of the National Academy of Sciences of Ukraine.

## REFERENCES

1. J.D. Vergados, H. Ejiri, F. Simkovic, *Rep. Prog. Phys*. **75**, 106301 (2012).
2. A. Giuliani, A. Poves, *Adv. High. En. Phys*. **2012**, 857016 (2012).
3. J.J. Gomez-Cadenas et al., *Riv. Nuovo Cim*. **35**, 29-98 (2012).
4. P. Belli et al., *Nucl. Phys. A* **846**, 143-156 (2010).
5. H.V. Klapdor-Kleingrothaus, I.V. Krivosheina, *Mod. Phys. Lett. A* **21**, 1547-1566 (2006).
6. J.E. Kim, G. Carosi, *Rev. Mod. Phys*. **82**, 557-601 (2010).
7. M. Laubenstein et al., *Appl. Radiat. Isotopes* **61**, 167-172 (2004);
   C. Arpesella, *Appl. Radiat. Isotopes* **47**, 991-996 (1996).
8. M. Berglund, M.E. Wieser, *Pure Appl. Chem*. **83**, 397-410 (2011).
9. M. Wang et al., *Chinese Phys. C* **36**, 1603-2014 (2012).
10. V.I. Tretyak, Yu.G. Zdesenko, *At. Data Nucl. Data Tables* **80**, 83-116 (2002).
11. E.B. Norman, *Phys. Rev. C* **31**, 1937-1940 (1985).
12. P. Belli et al., *Eur. Phys. J. A* **42**, 171-177 (2009).
13. P. Belli et al., *Phys. Rev. C* **87**, 034607 (2013).
14. E. Andreotti et al., *Appl. Radiat. Isotop*. **70**, 1985-1989 (2012).
15. P. Belli et al., *Nucl. Phys. A* **859**, 126-139 (2011).
16. M.I. Krivoruchenko et al., *Nucl. Phys. A* **859**, 140-171 (2011).
17. S.A. Eliseev et al., *J. Phys. G* **39**, 124003 (2012).
18. P. Belli et al., *Eur. Phys. J. A* **49**, 24 (2013).
19. P. Belli et al., *Eur. Phys. J. A* **47**, 91 (2011).
20. K. Blaum, J. Dilling, W. Nortershauser, *Phys. Scr. T* **152**, 014017 (2013).
21. P. Belli et al., *Phys. Rev. C* **83**, 034603 (2011).
22. P. Belli et al., *Nucl. Phys. A* **806**, 388-397 (2008).
23. P. Belli et al., *Phys. Lett. B* **711**, 41-45 (2012).